# Black-boxing and cause-effect power

William Marshall[1], Larissa Albantakis[1], Giulio Tononi[1, *]

[1]*Department of Psychiatry, Center for Sleep and Consciousness, University of Wisconsin, Madison, WI, USA*
*Corresponding author: gtononi@wisc.edu*

## Abstract

Reductionism assumes that causation in the physical world occurs at the micro level, excluding the emergence of macro-level causation. We challenge this reductionist assumption by employing a principled, well-defined measure of intrinsic cause-effect power – integrated information ($\Phi$), and showing that, according to this measure, it is possible for a macro level to "beat" the micro level. Simple systems were evaluated for $\Phi$ across different spatial and temporal scales by systematically considering all possible *black boxes*. These are macro elements that consist of one or more micro elements over one or more micro updates. Cause-effect power was evaluated based on the inputs and outputs of the black boxes, ignoring the internal micro elements that support their input-output function. We show how black-box elements can have more common inputs and outputs than the corresponding micro elements, revealing the emergence of high-order mechanisms and joint constraints that are not apparent at the micro level. As a consequence, a macro, black-box system can have higher $\Phi$ than its micro constituents by having more mechanisms (higher composition) that are more interconnected (higher integration). We also show that, for a given micro system, one can identify local maxima of $\Phi$ across several spatiotemporal scales. The framework is demonstrated on a simple biological system, the Boolean network model of the fission-yeast cell-cycle, for which we identify stable local maxima during the course of its simulated biological function. These local maxima correspond to macro levels of organization at which emergent cause-effect properties of physical systems come into focus, and provide a natural vantage point for scientific inquiries.

## Author Summary

We challenge the reductionist assumption by studying causal properties of physical systems across different spatiotemporal scales. The result is that – contrary to reductionist views – causal power can emerge at macro scales. Rather than relying on the traditional notion of coarse-grains (averages), we introduce the notion of functional black boxes that are defined based on their input-output relationship. Using a sequence of examples, our work demonstrates that black boxes are particularly well suited to capture the heterogeneous and specialized nature of components in biological systems. While the emergence of coarse-grained systems relies on increased specificity, black-boxing reveals the importance of structure and integration. Our framework is mathematically rigorous and fully general, hence applicable across many disciplines; it is particularly useful in objectively identifying informative perspectives on complex systems in the physical sciences.

# Introduction

Reductionist approaches in science usually assume that the optimal causal model of a physical system is at the finest possible scale. Coarser causal models are seen as convenient approximations due to limitations in measurement accuracy or computational power (Kim, 2000; Nagel and Hawkins, 1961). The reductionist view is based on the conjecture that the micro level of causal interaction is causally complete, leaving no room for additional causation at a macro level (Kim, 1993). The reductionist assumption is most obvious in fields such as particle physics (Nakamura et al., 2010), neuroscience (Markram, 2012), and nanotechnology (Bhushan and Marti, 2010), but it can also be found in the social sciences (Imai et al., 2011), where researchers endeavor to 'look inside the black box'.

A case has been made for the occurrence of genuine emergence at various macro levels (Ellis, 2011; Fodor, 1974), such as the emergence of mind above and beyond the individual neurons (or atoms) that constitute the brain (Tononi, 2008), and for the autonomy of the special sciences such as chemistry (Scerri and McIntyre, 1997), and biology (Dupré, 2009; Walker and Davies, 2013), above and beyond the underlying physics. However, arguments in favor of emergence have often been vague, or they have focused on the possibility that macro variables may have greater descriptive power than micro variables, rather than greater causal power (List and Menzies, 2009; Pfante et al., 2014; Wolpert et al., 2014).

Inspired by statistical physics, macro-level descriptions of a system are typically taken to be coarse-grainings, i.e. averages over micro elements and micro time steps. The reductionist assumption has been challenged by the introduction of explicit measures of cause-effect power, which were used to show that such coarse-grainings can indeed have greater cause-effect power at the macro level (Hoel et al., 2013, 2016). In simulated examples of simple logic gate systems, we coarse-grained (nearly) identical elements ('neurons') into groups ('neuronal groups') and averaged over their states. We demonstrated that, under certain conditions involving degeneracy and/or indeterminism, a macro-level system of coarse-grained elements can "beat" the micro-level system in terms of cause-effect power (Hoel et al., 2013, 2016).

However, moving beyond statistical physics to biology, the macro elements of interest cannot be obtained by coarse-graining, because they are constituted of heterogeneous micro elements that are often compartmentalized and have highly specific functions, which would be muddled by averaging (see Box 1). For example, take the neuron, considered as the fundamental unit in much of neuroscience. Clearly, a neuron cannot be represented by a coarse-grained macro element, because it is constituted of a great diversity of specific molecules, organized in highly specific and hierarchical ways, performing highly specific functions. Indeed, it is the very specificity of the internal micro elements that makes the reductionist assumption seem inevitable in these cases: while we can treat a neuron as a black box for ease of understanding and for convenience, it would seem that its full causal power can only be captured by considering all the molecules that constitute the black box, in exquisite and specific detail (Markram, 2006).

Here we further challenge the reductionist assumption by generalizing the causal analysis employed for coarse-graining to black-boxing (Tononi, 2010): we first analyze a system of heterogeneous, specific micro elements at the micro level; then we repeat the analysis at the macro level by grouping subsets of those micro elements inside black boxes (macro elements). Black boxes are characterized exclusively by their overall input-output function (Ashby and others, 1956; Bunge, 1963). The heterogeneous micro elements inside the black box are hidden inside a macro element, rather than averaged as with coarse-graining. As an example of a black box, Fig. 1 shows, on the left, a simple, schematic neuron constituted of a number of specific micro elements (synapses S, cell body C, and axon hillock A) that interact internally in specific ways. On the right, the neuron is treated as a single macro element, a black box, that receives inputs (spike or no spike for each input), produces a single output (spike or no spike), and conceals its micro elements inside.

In what follows, we assume that the causal power of a system is quantified by its intrinsic cause-effect power as previously defined (Oizumi et al., 2014; Tononi, 2015). While reductionism assumes implicitly that causal power resides exclusively with micro elements, we assess causal power explicitly – as intrinsic cause-effect power – and determine the spatiotemporal levels at which new cause-effect properties emerge. Such emergent cause-effect properties may include an increase in the overall intrinsic cause-effect power of the system, but also specific relationships between elements within the system ("mechanisms") that only become apparent at the macro level. To quantify intrinsic cause-effect power and system mechanisms at the micro level and all possible black-boxed macro levels, we use the interventional and counterfactual causal framework of integrated information theory (Oizumi et al., 2014; Tononi, 2015). As a measure of intrinsic cause-effect power, integrated information ($\Phi$) captures several aspects that are often overlooked in causal accounts (Oizumi et al., 2014): the dependence of cause-effect power on the specific state the system is in (state-dependency); how cause-effect power of the system is structured (composition); whether the whole system is causally irreducible to its parts (integration); and what defines the system's borders and grain (exclusion). These features make $\Phi$ particularly suited for assessing the cause-effect power intrinsic to a system, independent of external observers. As demonstrated through several examples, including the Boolean network model of the fission yeast cell cycle, the $\Phi$ value of systems of black-box macro elements can increase when going from finer to coarser spatiotemporal grains and lead to emergent cause-effect properties at macro scales.

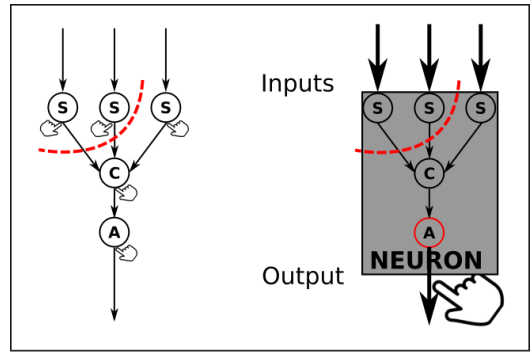

***Figure 1:*** *A schematic neuron considered as a number of `micro' elements (left), or as a black box (right). At the micro scale, the neuron receives inputs at its synapses (S), which are passed on to the cell body (C) and then to the axon hillock (A), which outputs to other neurons. Cause-effect power is assessed by perturbing each element (small hands) and observing the effects, while irreducibility is assessed by partitioning the elements (dashed red line). At the macro scale, there is only the black-box element (neuron) which receives three inputs and generates an output. Cause-effect power is assessed by perturbing the output of the black box (big hand) and observing its effects without constraining the constituent micro elements, however its irreducibility is still assessed by partitioning between micro elements (dashed red line).*

## Box 1 — Black-boxing and coarse-graining

A discrete, finite physical system can be considered at various spatiotemporal levels. At the most fine-grained scale, it is constituted of a set $S_m$ of micro elements, each having at least two states. Supervening, physical macro-level systems $S_M$ can be obtained by a mapping $M: S_m \rightarrow S_M$ that groups disjoint subsets of $S_m$ into non-overlapping macro elements. A physical macro element is thus constituted of one or more micro elements, operating over one or more micro time steps and can be manipulated, observed, and partitioned. For each macro element, M defines how the states of its constituting micro elements are mapped onto the possible states of the macro element. In previous work (Hoel et al., 2013, 2016), we demonstrated the emergence of cause-effect power in 'coarse-grained' macro-level systems with average-based state mappings. Here, we extend these results to 'black-box' macro elements with an output-based state mapping (Box Figure).

**Coarse-graining:** Coarse-graining corresponds to the notion of a macro state in statistical physics. In coarse-graining, the state mapping is a function that depends only on the average of the micro states of the micro elements constituting the macro element, without reference to the identity of individual micro elements (Hoel et al., 2013, 2016). This means that all micro states with the same average have to be mapped onto the same macro state, e.g., $s_m = \{00, 10, 01\} \rightarrow s_M = $ 'OFF' and $s_m = \{11\} \rightarrow s_M = $ 'ON', while $s_m = \{00, 10\} \rightarrow s_M = $ 'OFF' and $s_m = \{01, 11\} \rightarrow s_M = $ 'ON' would not be a proper coarse-grain mapping.

**Black-boxing:** Black boxes correspond to the typical notion of macro elements in the special sciences, such as cells or organisms in biology. In black-boxing, the state of a macro element is determined by the state of its output (micro) elements at a specific (micro) time step, without reference to the states of its internal micro elements. A possible mapping for the schematic system shown in Box Figure (left) in which 5 micro elements form a black box is, e.g., $s_m(t_3) = \{XXXX0\} \rightarrow s_M = $ 'OFF' and $s_m(t_3) = \{XXXX1\} \rightarrow s_M = $ 'ON'. This means that, given an input at time $t_0$, the macro state of the black box corresponds to the micro state of the output element at time $t_3$, while the states of the hidden elements are ignored.

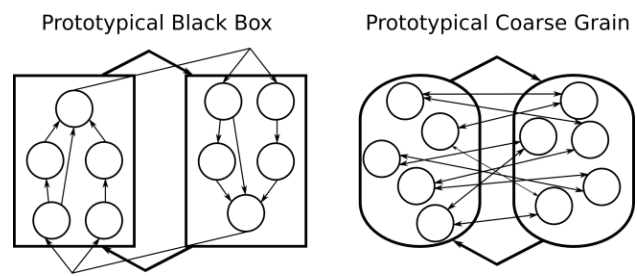

Prototypical Black Box          Prototypical Coarse Grain

***Box Figure:*** *Prototypical examples of black-box and coarse-grain elements that can lead to the emergence of macro-level cause-effect power. Left: A black-box element conceals many micro elements with specific functions. Right: A coarse grain macro element averages together many homogenous micro elements that share a global function.*

**Increasing intrinsic cause-effect power:** In recent work, we showed that coarse-grained physical systems can, under certain conditions, 'beat' the corresponding micro-level system in terms of measures of effectiveness (Hoel et al., 2013) and intrinsic cause-effect power (Φ) (Hoel et al., 2016). As done in this study, we simulated simple physical systems constituted, at the micro level, of collections of logic gates. The main factor enabling higher intrinsic cause-effect power through coarse-graining is a reduction in indeterminism and degeneracy at the macro level (Hoel et al., 2013, 2016). Determinism and degeneracy affect the *selectivity* of a system in its current state. In a non-degenerate and deterministic system, the current system state constrains with maximum selectivity both the cause repertoire (only one past state is possible - no degeneracy) and the effect repertoire (only one future state is possible – no indeterminism). In a degenerate system, multiple past states could lead to the current state of the system. In a non-deterministic system, multiple future states could follow the current state. Grouping noisy or degenerate micro elements into less degenerate and more deterministic macro elements may lead to a gain in the selectivity of the system's mechanisms. Everything else being equal, more selective mechanisms have higher intrinsic cause-effect power $\varphi$ (see Theory), which translates to higher $\Phi$ at the system level and thus may lead to emergence of macro-level cause-effect power in coarse-grained systems (Albantakis and Tononi, 2015; Hoel et al., 2016).

In general, coarse-graining micro systems, in the sense of averaging over subsets of them, may increase intrinsic cause-effect power when the constituting micro elements are all roughly of the same kind and all their inputs and outputs can be treated as equivalent. However, in system architectures constituted of heterogeneous micro elements, with highly specific functions, which are typical for biological and electronic systems, averaging across micro states may blur rather than enhance cause-effect power. It is these types of modular system architectures for which black-boxing is particularly suited to bring about emergent cause-effect properties at the macro level: in the results section, we demonstrate that black-boxing may reveal high-order macro mechanisms that are not present at a micro scale. In turn, these support a more integrated cause-effect structure and higher $\Phi$ values at the macro level.

Taken together, mapping a finer-grained system into a coarser, macro-level system may increase intrinsic cause-effect power both through coarse-graining (possible increase in selectivity) or black-boxing (possible increase in integration). Which mapping is more suited to bring about emergent cause-effect properties depends on the type of system architecture. Ultimately, we can consider a continuum of possible macro elements combining the two complementary approaches as the general case, where black boxes with one output for all micro elements of a box at a particular micro time-step and coarse grains with an output for each micro element are the extremes.

# Theory

Integrated information ($\Phi$) measures the intrinsic cause-effect power of a physical system (Albantakis and Tononi, 2015; Oizumi et al., 2014) by evaluating five requirements: the system's capacity to make a difference to itself (intrinsicality), composition, information, integration, and exclusion. Loosely defined, $\Phi$ quantifies to what extent a system's cause-effect structure, which specifies how all the system's parts constrain each other's past and future states, is integrated, that is, irreducible to subsystems (more below). The measure $\Phi$, which was developed as part of integrated information theory (IIT) (Oizumi et al., 2014), builds on interventionist, counterfactual accounts of causality (Lewis, 1974; Pearl, 2009) and can also be employed as a general measure of complexity that captures to what extent a system is both integrated and differentiated (Albantakis and Tononi, 2015; Albantakis et al., 2014; Hoel et al., 2016).

We formally define a physical system as a set of elements, for example neurons in the brain or logic gates in a computer, such that each element has at least two states, inputs that can influence these states, and outputs that in turn are influenced by these states. Furthermore, it must be possible to manipulate, observe, and partition among elements, in order to evaluate their cause-effect power. To fully characterize the cause-effect properties of a physical system, we first randomly perturb its elements into all possible states according to a maximum entropy distribution and observe their subsequent state transitions. Through this process, one obtains the transition probability matrix (TPM) for the physical system. During the perturbations, elements outside the physical system under consideration are held fixed; the states of these elements are considered "background conditions" (Oizumi et al., 2014). By fixing the background conditions we control external influences and use the system's TPM to calculate its intrinsic cause-effect properties, including $\Phi$ (see Supp. S1 Text).

Given the TPM of a system, the next step is to identify all its mechanisms—the subsets of the system which, in their current state, have irreducible cause-effect power within the system itself (intrinsicality). To this end, we test the entire power-set of system elements as candidate mechanisms (composition). To have irreducible cause-effect power, a set of elements in its current state must selectively constrain the potential past and future states of the system (information). This is evaluated using the conditional probability distribution of past or future states given the current state of the set of elements. A mechanism can be composed of one or more elements, as long as it constrains the past and future states of the system above and beyond its parts (integration). The degree to which a mechanism in its current state is irreducible is measured by $\varphi$, which quantifies the irreducible cause-effect power of the mechanism within the system (Albantakis and Tononi, 2015; Albantakis et al., 2014; Marshall et al., 2016; Oizumi et al., 2014; Tononi, 2015). In the following, we distinguish between mechanisms consisting of a single element (first-order mechanisms) and those composed of multiple elements (high-order mechanisms), which play an essential role in integrating the whole system. Note that a set of elements that fails to irreducibly constrain the system's past state does not have any potential causes within the system, and a set of elements that fails to constrain the system's future state irreducibly does not have any potential effects within the system; in both cases $\varphi = 0$ and neither is an intrinsic mechanism of the system.

The set of all mechanisms within a system defines its cause-effect structure. If a candidate mechanism in its current state has a value of $\varphi = 0$, then it is reducible, and does not contribute to the cause-effect structure of the system (see Fig. 4). The intrinsic cause-effect power of the system is quantified by its integrated information $\Phi$ (Albantakis and Tononi, 2015; Albantakis et al., 2014; Marshall et al., 2016; Oizumi et al., 2014; Tononi, 2015), which captures the irreducibility of the cause-effect structure: the degree to which the system's cause-effect structure is changed by partitioning the system (eliminating constraints among parts). For $\Phi$ to be high, every possible partition must affect many mechanisms that constrain the system in a highly selective, irreducible manner (having high $\varphi$). If $\Phi = 0$, then there is at least one part of the system that remains unconstrained by the mechanisms of the rest: from the intrinsic perspective, there is no unified system, even though an external observer can treat it as one.

Finally, from the intrinsic perspective, the set of elements that form a system must be definite. In other words, it must have a self-defined causal border with its environment that identifies the elements within the border as part of the system, while elements outside the border belong to the system's environment. Even though many subsets and supersets of elements may have $\Phi > 0$, only sets of elements that specify a local maximum of $\Phi$ have well defined borders from the intrinsic perspective (exclusion). A system's border is thus defined by the intrinsic cause-effect structure of its elements, such that adding or removing a single element will result in a decrease of cause-effect power.

This exclusion principle also applies across spatiotemporal scales: from the intrinsic perspective, the set of elements that form a system must have a definite spatiotemporal grain. As with the system's borders, it is the intrinsic cause-effect structure that self-defines its spatiotemporal scale, which is one that is a local maximum of $\Phi$. Local maxima of $\Phi$ identify those scales at which cause-effect properties emerge – any finer or coarser grains necessarily result in a reduction of cause-effect power and a blurring of intrinsic cause-effect properties. To evaluate intrinsic cause-effect power at macro scales and identify the definite scales at which new cause-effect properties emerge, micro elements can be grouped either by coarse-graining as in Hoel et al., (2013, 2016) or, more generally, by black-boxing, as will be demonstrated here.

## Black-boxing

In typical usage, a black box is an object into which inputs impinge and from which outputs emerge, but its internal workings are not available for inspection (Ashby, 1956; Bunge, 1963). For our purposes, a 'black-box element' is a physical macro element that can be manipulated, observed, and partitioned, which is constituted of several micro elements (spatial), operating over several micro time steps (temporal). To qualify as a black box, it must satisfy the following conditions:

(i)  It must have at least one input, one output, and two or more (macro) states that can be read from its output (element)

(ii)  The micro elements and micro updates within the black box are hidden (black box condition)

(iii) The micro elements contribute causally to the black box's output (integration)

(iv)  There cannot be any overlap between the micro elements of multiple black boxes (exclusion)

Specifically:

(i)   The inputs and outputs of a black box are defined in terms of the internal micro elements that receive direct input from other elements/black boxes (e.g., synapses S in Fig. 1) and directly output to other elements/black boxes (e.g., the axon hillock A in Fig. 1). For this work we allow for inputs to arrive at multiple micro elements, but restrict outputs to leave from only a single micro element within the black box. Furthermore, the inputs are taken to arrive at the beginning of the macro time step, while the outputs are taken to depart at the end of the macro time step. In principle, this framework could be extended to multiple output elements and to a more general treatment of time steps by allowing macro elements with different temporal grains.

(ii)  The state of a black-box element is taken to be the state of its (micro) output element at its (micro) output time step. The transition probabilities associated with a black-box element are determined as usual by causal analysis, perturbing the inputs of the black box into all possible states according to a maximum entropy distribution. At the end of the macro update, the state of the black box is observed from its output element (see Fig. 2). In this way, one can determine the cause-effect power that the inputs (i.e., outputs from other black-box elements) have on the state of the black-box element over the respective macro update. In line with the notion of black boxes, the micro elements within the black box are "hidden" from other black boxes within the system, meaning they do not directly contribute to the intrinsic cause-effect power of the system, but only indirectly through their black box's output. Any other direct micro interactions are not considered intrinsic to the macro-level system and therefore do not contribute to its cause-effect power at all (see Supp. S3 Text). Crucially, for the duration of the macro update, the internal elements are allowed to evolve unperturbed; however, to discount the cause-effect power of micro elements when evaluating $\Phi$, the initial states of micro elements and any micro connections leaving the black box, other than its designated output element at the designated output time step, are noised during the perturbation analysis. A consequence of this perturbation procedure is that potential causes and effects must be direct (i.e. between two black boxes), and that potential causes and effects that are mediated by a third black box are 'screened off' and do not contribute to cause-effect power (Supp. S3 Text, Fig. S4).

(iii) The requirement that every constituent micro element must causally contribute to the output of its black box is mandated by the integration principle that cause-effect power must be irreducible. Even at the macro level, a system can only be integrated if its micro level is integrated. Moreover, it is not meaningful to consider a black-box element as a single physical element if it is reducible to two or more unrelated elements. The requirement of micro integration is satisfied implicitly when assessing models using integrated information; any physical system that violates it will be found to be reducible and thus have $\Phi = 0$, as even for macro systems, $\Phi$ is evaluated by partitioning between micro elements. This implies that it is not possible to take a non-integrated system of micro elements and to black-box it in such a way as to create an integrated system of macro elements (Supp. S3 Text, Fig S5).

(iv) The requirement for no overlap among the constituents of different black boxes (or equivalently that a micro element cannot be a constituent of more than one black-box element) is a consequence of causal exclusion. A physical (macro) element must be definite, meaning that it has a well-defined border which separates it from other macro elements. The importance of the exclusion condition has been independently recognized in the theory of computation: it is only meaningful to say that a physical system implements a computation if the system is constituted of distinct, non-overlapping elements (Chalmers, 1996). If black-box elements were permitted to overlap, then every open physical system could be said to implement any computation (Chalmers, 1996; Putnam and Putnam, 1988).

Together, the above requirements allow to specify inputs and outputs of each black-box element, to define its macro state, to include within each black box only micro elements that are integrated and contribute to its input-output function, and to draw 'borders' around each black-box element that exclude any overlap with other black boxes (Fig. S6).

**Local Maxima of cause-effect power**

Only systems that support local maxima of $\Phi$, both in terms of constitution and spatiotemporal grain, are definite and have intrinsic cause-effect power. A system of elements is a local maximum if there are no 'neighboring' systems with a higher value of $\Phi$. When only micro elements are considered, such as in (Marshall et al., 2017), it is natural to define a neighbor as any system that differs in constitution by only a single micro element, that is, any system that can be made by either adding or removing a single element. However, to determine whether two systems at different spatiotemporal grains are neighbors, several distance measures have to be taken into account. For the present purposes, we consider three different distances between systems to establish whether two systems are neighbors in this general context. The first is the constitutional distance between two systems, which is the number of micro elements that must be added / removed from one system to transform it into the other. Next is the temporal distance between two systems, which is the difference in the number of micro updates that make up the corresponding macro updates. Finally, the spatial distance between two systems is the distance between the partitions that group micro elements into macro elements. In the current work we use the maximum matching distance between partitions (Almudevar and Field, 1999), which is essentially the number of micro elements that must be moved from one grouping to another. If the sum of the constitutional, temporal and spatial distances between two systems is equal to 1 then those systems are neighbors, i.e., two systems are neighbors if they differ by a single step in exactly one of the three distances.

Given a set of micro elements, we evaluate all possible systems (sets of black-box elements) to determine which systems have intrinsic cause-effect power, at which spatiotemporal grain (the set of black-box elements that define the system), and what their borders are (the set of micro elements that constitute the system). Evaluating all possible sets of black-box elements includes all possible groupings of micro elements into macro elements. Then, for every grouping all possible elements of each black box are considered as its output element. Finally, cause-effect power is evaluated over all possible macro time steps of each black-box system. Note that not all micro elements must be grouped into black boxes when searching for maxima of intrinsic cause-effect power. It may be that adding a specific micro element to any black-box element within the system would in fact reduce cause-effect power. In this case, such micro elements are held fixed as background conditions of the macro system (see Supp. S3 Text).

# Results

In the following, we demonstrate black-boxing and its importance for revealing macro-level cause-effect properties based on a set of simple proof-of-principle examples before we apply the framework to a biological model of the fission-yeast cell-cycle. Crucially, we demonstrate that systems of black-box macro elements can have higher intrinsic cause-effect power than their corresponding micro systems, and support local maxima of $\Phi$ that reveal emergent functional properties. For the purposes of this work, we shall consider collections of elements that are binary micro elements which cannot be further reduced or split, and the time scale of state transitions to be a micro time step. Time is implicit in the TPM, as micro elements are synchronously updated at discrete micro time steps. In principle, integrated information is defined for any discrete system of elements. The full mathematical details of the $\Phi$ calculation are described elsewhere, we recommend (Oizumi et al., 2014) but details are also available in (Hoel et al., 2016; Marshall et al., 2016; Tononi, 2015); full example analyses are presented in Supp. S1. All calculations in this work were performed using the PyPhi software package in Python (Mayner et al., 2016), which includes a documented example for a black-box analysis.

## How macro beats micro: composition and integration

An intuitive example in which black-boxing may be appropriate is propagation delay – the amount of time between the output of one element and its effect on another element. Such delays are largely ignored in functional analyses and are taken to be an implicit aspect of the element of interest, *i.e.*, they are black-boxed. In the context of logic gates, for example, NOR logic is commonly described as a "universal" in the sense that any other logic can be built strictly from NOR gates. However, building, say, an XOR gate from NOR gates requires in fact a propagation delay as an implicit part of the circuit.

In the following example, we explicitly model such propagation delays as (one or more) COPY elements that take a single input and then output the same value. Fig. 2 shows the micro structure of an XOR element with a one-step propagation delay, along with the corresponding macro element, a black box with XOR logic.

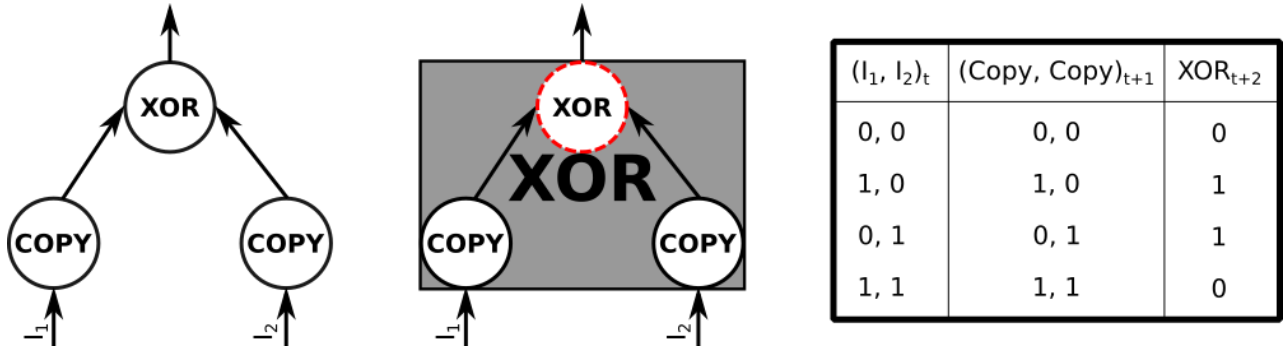

*Figure 2*: *An XOR logic gate with one time-step propagation delay. Left: Two COPY elements which each take a single input and relay the values to an XOR element. Middle: A black-box element with two inputs and a single output element (dashed red outline). Right: By perturbing the inputs to the black-box element in all possible ways, it is determined that it implements XOR logic over two time steps.*

Consider a system of three interconnected XOR elements with a one-step propagation delay. At the micro level, this system is constituted of nine micro elements—six COPY and three XOR, which can be black-boxed over two time steps into a macro system of three interconnected XOR elements (see Fig. 3). The current state of all elements is OFF.

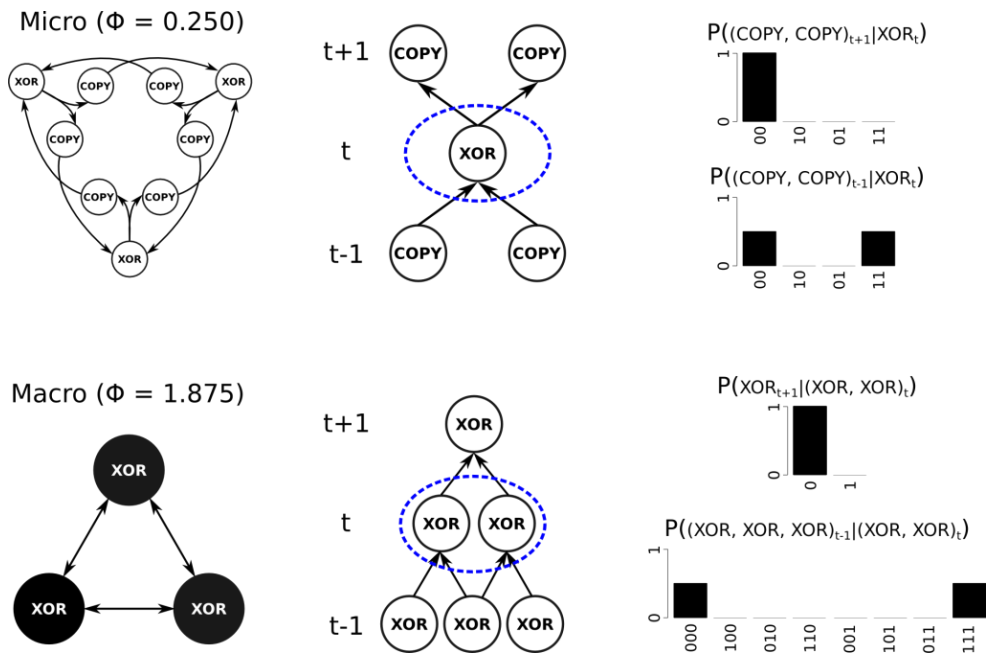

**Figure 3:** *A system of three interconnected XOR elements with one-step propagation delay and all elements in the OFF state. Top: At the micro level, the propagation delay is modeled by COPY elements between the XOR elements (left). The micro level has only first-order mechanisms specified by individual XOR elements (right), and Φ = 0.25. Bottom: When the elements are black-boxed over two time steps, the system is comprised of three interconnected macro elements implementing XOR logic. The macro-level system has second-order mechanisms specified by pairs of XOR elements (right) and Φ = 1.875.*

Assessing the cause-effect structure of the micro system, we find that there are only three first-order mechanisms and no high-order mechanisms. The three XOR elements each specify a mechanism with φ = 0.5: by being in the OFF state, each XOR specifies that its two inputs must have been either (OFF, OFF) or (ON, ON) and that its outputs, the COPY elements, must be OFF in the future (Fig. 3, top-right). All other sets of elements do not have cause-effect power, or are reducible, so φ = 0 (see Fig. 4). Recall that from the intrinsic perspective, a set of elements must constrain both the system's past and future irreducibly to be a mechanism for the system (see Theory). The six COPY elements, taken individually, lack any potential effect within the system: by being in the OFF state, a COPY by itself does not constrain the future state of its XOR output, which is still equally likely to be ON or OFF depending on the state of its other input (Fig. 4, top). On the other hand, two COPY elements in the state (OFF, OFF) that input to the same XOR element do irreducibly constrain the system's future states, since together they specify that the XOR element they output to will be OFF. Nonetheless, these pairs of COPY elements do not form a second-order mechanism in the system since their constraint on the system's past state is reducible: in the OFF state, the two COPY elements taken individually already specify that their inputs must have been OFF, leaving no room for additional second-order constraints (Fig. 4, bottom). The lack of either irreducible past or future constraints thus prevents the COPY elements from specifying first- or high-order mechanisms in the system. The integrated information of the micro physical system is $\Phi = 0.25$ (see Supp. S1 Text).

The macro-level physical system with black-box elements also has three mechanisms with $\varphi = 0.5$, but they are second-order mechanisms specified by pairs of XOR elements. By being in the state (OFF, OFF), each pair of XOR elements specifies that the past state of the entire model must have been either (OFF, OFF, OFF) or (ON, ON, ON), and that the future state of their common output must be OFF (Fig. 3, bottom-right). Neither of the XOR elements in this high-order mechanism can specify these constraints on its own. Individual XOR elements lack potential effects in the system for the same reason as the individual micro COPY gates above. At the macro level, the collection of mechanisms (cause-effect structure) is more integrated than that of the micro level, with a value of $\Phi = 1.875$. Although the system has the same number of mechanisms and the same $\varphi$ values at both the micro and the macro level, the black-boxed system has higher $\Phi$ because a system partition impacts the macro level cause-effect structure more than the micro level cause-effect structure. The black-box system "wins" by having more overlap in its mechanisms, both in terms of the elements they are composed of and the constraints they impose. The high-order mechanisms of the black-box system have overlapping constraints, with each mechanism constraining all elements within the system, whereas the first-order mechanisms of the micro system only constrain their respective COPY inputs and outputs, without overlap. A system partition at the micro level thus only affects a single micro mechanism, whereas a system partition at the black-box level affects all of the mechanisms in the system, resulting in higher integration (see Supp. S1 Text). Consequently, there is irreducible cause-effect power that emerges at this macro level of the physical system. Concealing the COPY elements inside the black boxes reveals the high-order interactions between the XOR gates over two time steps. Note also, that, while the causal analysis is state-dependent, in this example the irreducibility of micro and black-box cause-effect structures (their $\Phi$ values) and thus the relationship between levels, is equivalent for all possible system states.

**Figure 4:** *Two potential mechanisms from the propagation delay network in Fig. 3 which end up being reducible. On top is a COPY element that does not specify a mechanism. By being OFF in the current state, the COPY element constrains its input to be OFF in the previous state, but it does not constrain the future state of its output element, because the state of the XOR element still completely depends on the unknown state of its other input (shown here in grey). The bottom panel is a set of COPY elements which do not specify a high-order mechanism because they do not have an irreducible cause (the red line partitions the cause in two with no loss of information). Taking each COPY element independently fully constrains the past state of its input to be OFF.*

## Finding local maxima of intrinsic cause-effect power

In a second example, we consider a larger micro system constituted of 55 elements that all implement NOR logic. By testing all possible black-boxings, we establish three local maxima of cause-effect power which reveal the organizational hierarchy of the system. Fig. 5, demonstrates how a group of 11 elements implementing NOR logic can be connected in such a way to produce AND/OR logic, or MAJORITY logic at coarser spatiotemporal scales.

The 55-element system is arranged into five interconnected groups of 11 elements, with each group organized according to Fig. 5 so that the system exhibits different functions at different spatiotemporal scales. Each group of 11 elements receives inputs from three other groups and has a single element that outputs to three other groups (Fig. 6, top left). We consider the system state in which each of the 55 NOR micro elements is ON. In the following, we focus on the cause-effect structures of the system levels shown in Fig. 5: the micro physical system of NOR elements, a black-boxed system of AND/OR elements, and a black-boxed system of MAJORITY elements. These systems are shown in Fig. 6 (top row) ordered according to the average spatial grain of their elements. Many other possible black-boxing schemes were also evaluated; however none of these other black-box macro systems had $\Phi$ > 0. Two representative examples of macro systems with $\Phi$ = 0 are included in Fig. 6 (bottom).

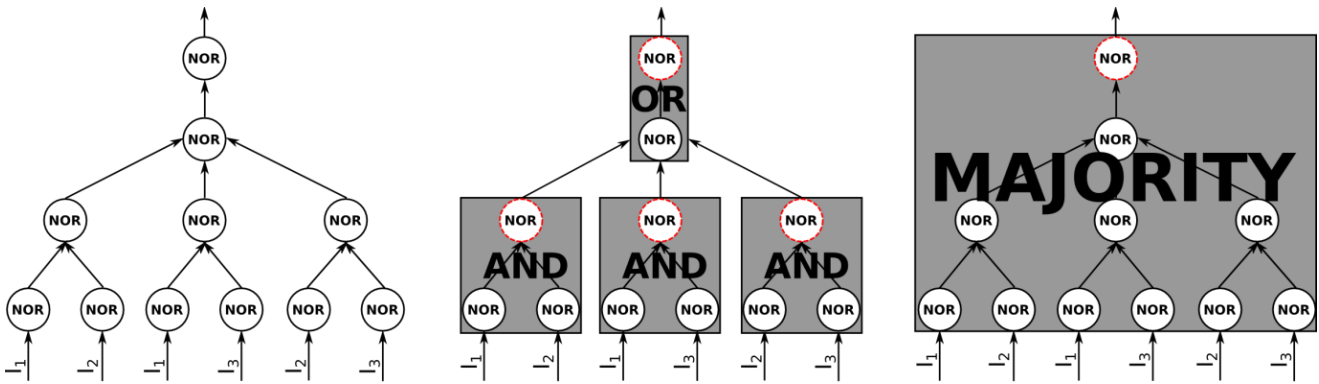

*Figure 5:* *Left: A collection of 11 NOR elements at the micro scale. Middle: A macro scale black-boxing of these elements into four black boxes. Perturbing the inputs of these black boxes reveals that three implement AND logic and the final one OR logic, each at a time scale of two time steps. Right: A macro scale black-boxing with one element, implementing MAJORITY logic over its three inputs at a time scale of four time steps. Note that there are only three inputs, but each input arrives at two different micro elements.*

At the micro level, the system's cause-effect structure consists of 55 first order mechanisms, one for each micro element with φ = 0.239 on average, and no high-order mechanisms. The integrated information of this micro physical system is Φ = 0.453 (see Supp. S1 Text).

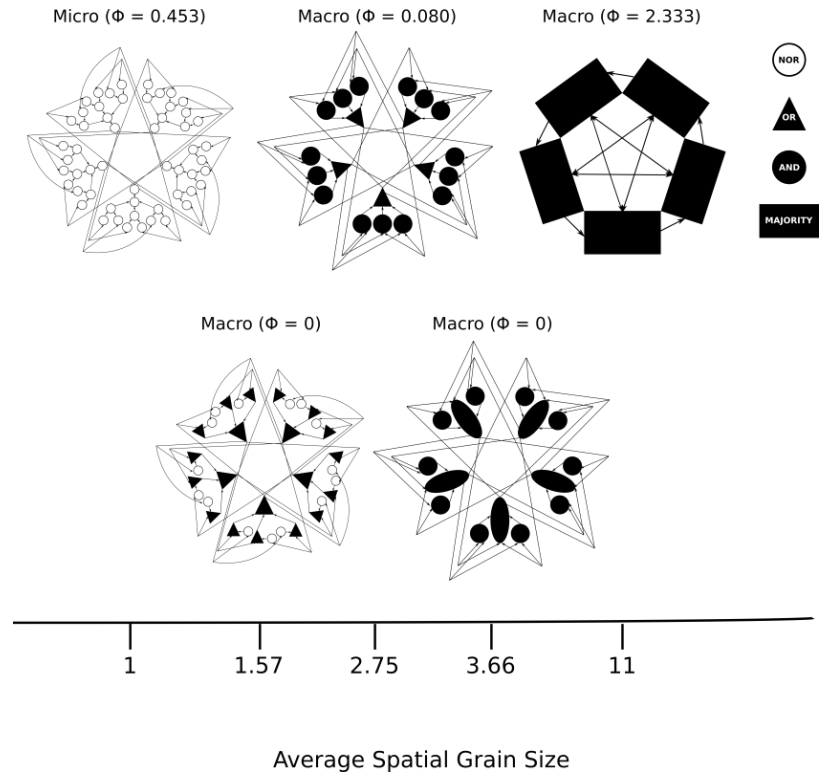

*Figure 6:* *A system of 55 interconnected NOR micro elements viewed at several different grain sizes, with all elements in the ON state. The micro elements form 5 interconnected groups, with each group arranged according to Fig. 5. The systems are arranged with micro elements on the far left and black-box elements of increasing spatial grain to the right. The legend on the right specifies the input-output function of each element. Each of the three systems on the top row is a local maximum of cause-effect power, corresponding to the three spatiotemporal grains shown in Fig. 5. Shown in the bottom row are two representative examples of the many systems with Φ = 0 at spatial grains between the local maxima.*

The macro-level AND/OR black-boxed system with an average spatial grain of 2.75 (Fig. 6, top, middle) has 20 macro elements, 15 implementing AND logic and 5 implementing OR logic, operating over two time steps. Similar to the micro level, its cause-effect structure is composed of 20 first order mechanisms (one for each black-box element) but no high-order mechanisms, with $\varphi = 0.112$ on average. This black-boxing reduces the number of first-order mechanisms, but does not reveal high-order mechanisms or overlapping constraints, thus the macro system is no more integrated than the micro system. Moreover, this black-boxing in fact reduces the integrated information of the first-order mechanisms in the system compared to the micro level ($\varphi$ values are 0.127 lower on average), leading to lower integrated information for the system ($\Phi = 0.080$).

The macro-level black-boxed system with an average spatial grain of 11 (Fig. 6, top right) is defined by considering black-box elements implementing MAJORITY logic over four time steps. Compared to the macro level with an average spatial grain of 2.75, this additional black-boxing step further reduces the number of elements, but increases the average $\varphi$ to 0.216 ($\varphi$ values are still 0.023 lower than the micro level on average). However, this macro system is endowed not only with first-order mechanisms, but with all possible second, third and fourth-order mechanisms. In total, its cause-effect structure includes 30 of 31 possible mechanisms from the power set of black-box elements, resulting in high integration, with $\Phi = 2.333$, more than the micro level.

Fig. 6 also shows additional black-box systems with $\Phi = 0$. One of these black-box systems with an average spatial grain of 1.57 has 20 black-box OR elements over two time steps and 15 micro NOR elements. A second black-box system with average spatial grain of 3.66 has 10 black-box AND elements over two time steps and 5 black-box AND elements over four time steps. For both of these systems (and many others not shown), the integrated information is $\Phi = 0$, because there is no common temporal scale over which all the elements in the system have effects on other elements within the system. For any specific temporal scale, there will be elements that do not causally contribute, thus the system is not integrated.

In summary, this example demonstrates how evaluating cause-effect power over many different spatial and temporal scales of black boxes identifies local maxima of cause-effect power and reveals emergent cause-effect properties. For this example, the analysis reveals functional relationships between elements; local maxima of cause-effect power occur specifically at the micro level of NOR elements (average spatial grain size of 1, $\Phi = 0.453$), at an intermediate macro level of AND/OR elements (average spatial grain size of 2.75, $\Phi = 0.080$) and at a coarser macro level of MAJORITY elements (average spatial grain of 11, $\Phi = 2.333$). While these spatial grains reveal emergent levels of organization at which the system exhibits intrinsic cause-effect power, which shed light on its cause-effect properties, the vast majority of systems of black-box elements, on the other hand, yield $\Phi = 0$.

## Boolean network model of the fission yeast cell cycle

As a demonstration of black-boxing in biological systems, we apply the framework to the Boolean network model of the fission-yeast cell-cycle (Davidich and Bornholdt, 2008). The model consists of nine Boolean ("micro") elements representing the state of crucial proteins expressed during cell division. Each element implements linear threshold logic, and the connections between elements are weighted, with each connection being either excitatory (+1) or inhibitory (-1) in nature (see Fig 7A). One element, "SK" only inputs to the system, receiving no feedback. This element acts as a catalyst for cell division: when it is activated while the network is in its biological attractor state, the remaining eight elements cycle through a sequence of 9 states, eventually returning to the initial attractor state (see Fig 7B). This cycle of states is called the 'biological sequence' of the model, and captures the specific sequence of protein expressions that occur during the cell-division cycle.

Since the element SK receives no feedback from the rest of the cell-cycle network, any system that includes SK will necessarily be reducible ($\Phi = 0$). Only when SK is fixed as a background condition can we potentially identify systems with $\Phi > 0$. Furthermore, if we consider the remaining eight elements (excluding SK) as a system, one of the states of the biological sequence ($t_2$, see Fig 7B) has no cause (potential past state) within the system (it is caused by the catalyst element SK which initializes cell division from outside the system). For this reason, the cause-effect structure of this system is undefined in state $t_2$. In what follows, we refer to the cell-cycle network as the eight strongly connected elements that contain both inputs and outputs (not including SK) and its biological sequence as the eight states ($t_1$, $t_3$-$t_9$) with well-defined cause-effect structures.

Previous work analyzing the cause-effect structure of the cell-cycle model demonstrated that the cell-cycle network constitutes a stable local maximum of integrated information across all states of the biological sequence (Marshall et al., 2017). However, this previous work only analyzed the cell-cycle model at the *micro* level, considering all possible subsets of micro elements. In the current work, we extend this analysis by considering the cell-cycle network at *macro* spatiotemporal scales. Specifically, we consider all possible groupings of the cell-cycle network into black-box macro elements, at time scales of 2, 3 and 4 micro updates (greater time scales may reveal additional local maxima and emergent cause-effect properties).

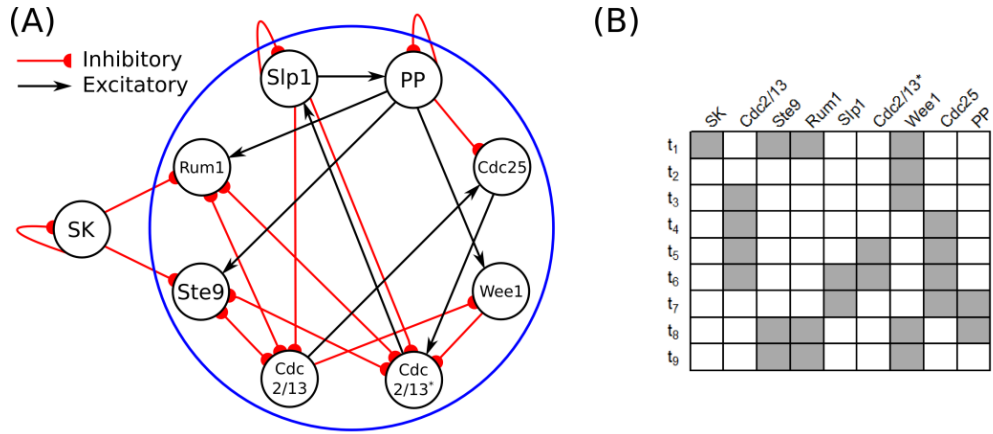

***Figure 7:*** *Boolean network model of the fission-yeast cell-cycle. (A) A network of 9 linear threshold elements connected by excitatory (black) and inhibitory (red) connections. The element, SK receives no input from the other elements. For this reason, we do not consider it as a part of the cell-cycle network, but rather as an external input that serves as a catalyst to initiate cell division. At the micro level, the system comprised of eight elements (in blue) is a stable local maximum of Φ for the duration of the biological sequence. (B) A sequence of 9 states that represent the cell division cycle, called the networks biological sequence.*

There are 4140 ways to group the eight micro elements in the cell-cycle network into any number of black-box elements, and for each grouping there are on average 10 different ways to define the output elements of the black boxes. Considering three different time scales for each set of black-box elements, results in a total of 124,176 macro systems to analyze. Across all states of the biological sequence, there are 2224 macro systems with $\Phi > 0$, an average of 278 per state, or roughly 0.22% of all possible systems.

Among the 2224 macro system with $\Phi > 0$, we identify 33 unique local maxima (some others are duplicates due to symmetries in the network). The majority of these local maxima are transient, occurring in an average of 2.5 out of 8 states in the biological sequence. However, 5 of the local maxima are stable over all states of the biological sequence. The micro system is one example of a stable local maximum, confirming that the results of (Marshall et al., 2017) hold even when considering macro systems. The remaining four local maxima occur at macro spatiotemporal scales, one at a time scale of 3 micro updates, and the others at a time scale of four micro updates (see Fig 8). Note that the intrinsic cause-effect power of a system is state dependent, and stability across subsequent time steps is not assumed at any point in the analysis. That the cell-cycle supports stable local maxima of macro cause-effect power is a feature of this biological system that is revealed by the causal analysis, rather than a requirement imposed by the framework.

Our analysis moreover reveals that one element in particular (Slp1) serves as a black box's output in every stable local maximum. This indicates that Slp1 may play a crucial role in stabilizing and integrating the network over longer time scales during the process of cell division—a property that could not be identified from its micro level interactions (Kim et al., 2013; Marshall et al., 2017).

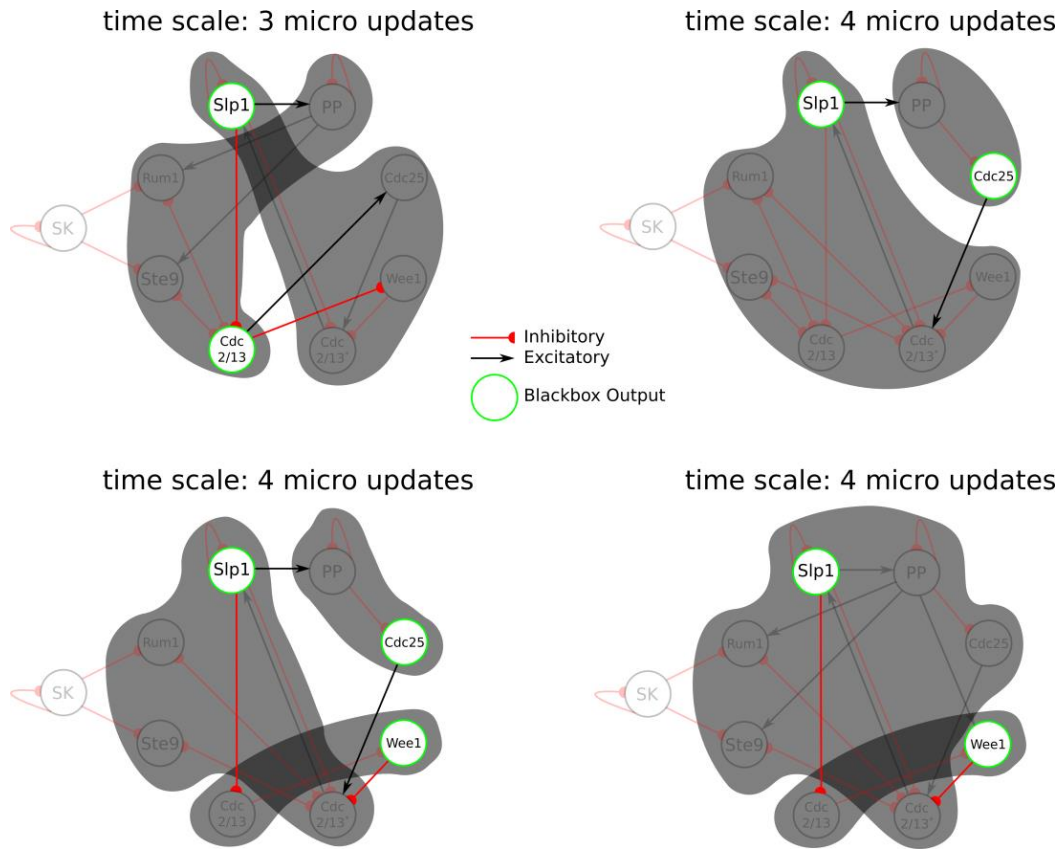

***Figure 8:*** *All stable local maxima of macro cause-effect power for the cell-cycle network over the course of its biological sequence. Stable local maxima are identified at two different time scales (over 3 or 4 micro updates) and with groupings of the eight micro elements into either two or three macro elements. The output element for each black box is marked by a green outline; one common feature among all of the stable maxima is that element Slp1 acts as an output element of one black box. Note that connections between black boxes that do not originate from output elements are not shown in the figure because they do not contribute to the cause-effect structure (see Supp. S3 Text).*

## Discussion

In this work we expand the framework for evaluating the cause-effect power of physical systems at multiple spatiotemporal scales, to include biologically motivated black-box macro elements defined by their input-output function. We then use this framework to explore the cause-effect power of simple systems of elements considered both at the micro level and after black-boxing, at a macro level. The cause-effect power of these systems was assessed using integrated information ($\Phi$), a measure of the cause-effect power that is intrinsic to a physical system. To properly capture cause-effect power from the intrinsic perspective of the system itself, $\Phi$ considers composition, specificity, irreducibility, and exclusion (Albantakis and Tononi, 2015; Oizumi et al., 2014). We show how macro systems based on black boxes can have higher intrinsic cause-effect power than any neighboring systems (including in some cases their micro element counterparts). This result complements and extends previous work that showed how intrinsic cause-effect power can increase when macro elements are defined by coarse-graining micro elements (Hoel et al., 2016). While coarse-graining may reduce degeneracy and/or indeterminism in a system, black-boxing may increase a system's intrinsic cause-effect power by increasing its integration.

Reductionist accounts of causation assume that all causal power resides with micro elements and time steps, excluding all macro levels (Kim, 2000). We argue that reductionist accounts of causation conflate the necessity of micro elements as constituents with their cause-effect power within the system. As shown in Fig. 4, a single micro element within a system may completely lack the power to constrain the system's future states—taken individually, it does not make any difference to the system. Yet, the high-order mechanism with irreducible cause-effect power shown in Fig. 3 would not exist without the individual micro elements to support it. Thus micro elements may play a role as a constituent of a high-order mechanism or a macro element with cause-effect power. The current work reveals the possibility that causal power may emerge at macro spatiotemporal scales, requiring only that a system is definite, with self-defined borders and spatiotemporal grain (by being a local maximum of $\Phi$). In such a case, the micro elements support the macro level as constituents, the macro level still supervenes upon the micro level, yet there are cause-effect properties that are only revealed at this particular macro level.

## Limitations and Future Work

In the current work, we use intrinsic cause-effect power as a quantification of causal power, and demonstrate several examples of systems of black-box macro elements with higher intrinsic cause-effect power than the corresponding micro systems. To the extent that the notion of causal power is appropriately captured and quantified by intrinsic cause-effect power, our results refute the reductionist assumption that causal power resides exclusively at the micro level. The value of our characterization of cause-effect power had been previously demonstrated in a number of contexts (Albantakis and Tononi, 2015; Albantakis et al., 2014; Marshall et al., 2017), and will continue to be evaluated in the future.

A limitation on the practical application of this framework is the computational demands for exhaustively evaluating intrinsic cause-effect power. Currently, cause-effect properties can only be fully explored for very small systems (< 10 micro elements; propagation delay example, cell-cycle example) or by exploiting symmetries in the system (local maxima example). Future work will extend the PyPhi software for evaluating intrinsic cause-effect power (Mayner et al., 2016) by including, for example, approximations based on the connectivity matrix. However, practical applications inevitably will have to use a targeted approach and only assess the intrinsic cause-effect power of a predetermined set of macro-level systems instead of evaluating all possible black-box systems. Theoretical investigations like the current work (see below) as well as previous exploration of coarse-grained macro elements (Balduzzi, 2011; Hoel et al., 2016) will be crucial to define the criteria that will guide such a targeted approach.

### Black-boxing reveals high-order mechanisms and joint constraints

The two main requirements for high $\Phi$ are that a physical system is differentiated (many specific mechanisms) and integrated (mechanisms with overlapping constraints). Typically, whenever a lower level system is mapped into a higher macro system, there is reduced state differentiation, i.e., the macro system has fewer elements and a smaller state space. This decrease in differentiation means fewer potential mechanisms and thus less potential integrated information (Marshall et al., 2016). In order for a macro level system to have higher cause-effect power ($\Phi$) than a finer grained system over the same elements, the macro system must increase cause-effect power either by having more specific mechanisms, or a more integrated set of mechanisms.

Degeneracy and indeterminism are two factors that influence the specificity of a mechanism. Everything else being equal, decreasing degeneracy and indeterminism leads to an increase in the cause-effect power of mechanisms within the system. In (Hoel et al., 2013, 2016) we demonstrated that coarse-graining (averaging) micro elements into macro elements can lead to an increase in intrinsic cause-effect power that can overcome the inherent loss of differentiation in macro systems. An increase in intrinsic cause-effect power through reduction of degeneracy is also possible through black-boxing, as shown in Supp. S2 Text.

The particular asset of black-boxing is that it may reveal high-order mechanisms and joint constraints between mechanisms at macro spatiotemporal scales. As demonstrated by the propagation delay example, the macro can even beat the micro level through increased integration. This may occur when elements with few potential effects are concealed within black-box elements, and micro elements with many potential effects serve as the outputs of black-box elements, resulting in a more densely interconnected set of macro elements, where groups of macro elements share common inputs and common outputs. If creating common inputs and common outputs among elements leads to additional, joint constraints on the possible past and future system states, elements may form high-order mechanisms, resulting in a more integrated cause-effect structure and higher $\Phi$. Being a part of high-order mechanisms, or being constrained by multiple mechanisms, gives an element additional ways to contribute to the cause-effect structure; when an element contributes in multiple ways, cutting that element has a greater effect on the cause-effect structure, making the system more irreducible. Being more irreducible means having higher intrinsic cause-effect power ($\Phi$) and may thus lead to a causally emerging macro level. This suggests that black-boxing is most beneficial when there are "causal bottlenecks" in the micro system, that is, when a micro element with a single or few outputs connects to a micro element with a single or few inputs. In such cases, it is impossible for these micro elements to contribute to high-order mechanisms, and such elements represent a "weak link" in the integration of the system. More generally, black-boxing should be particularly appropriate in systems with local modular interactions whose results are distributed across the system, such as molecular interactions within neurons in the brain, or electrical interactions within computer networks.

## Local maxima of intrinsic cause-effect power

Evaluating cause-effect power of black-box systems across many spatiotemporal scales shows that, in general, there can be several local maxima of macro cause-effect power, between which integrated information decreases or falls to zero. In Fig. 6, the local maxima capture emergent functional roles of black-box macro elements, corresponding to the different descriptions of the system as sets of NOR, OR/AND, or MAJORITY elements. Importantly, even within a given spatiotemporal grain, there will generally be several local maxima corresponding to overlapping subsets of elements, such that adding or subtracting an element reduces integrated information (Hoel et al., 2016; Oizumi et al., 2014). These local maxima of intrinsic cause-effect power across and within levels correspond to organizational macro levels and systems having emergent cause-effect properties. These are natural levels and systems for the special sciences to investigate.

A prime example is biological systems, since they contain many highly specialized components which are required to perform their function. In biology we can study the molecules within an individual cell, the interactions between networks of cells (nervous system), individual organs (liver, kidneys), whole organism (animals, humans), and communities of organisms (swarms, societies). The Boolean network model of the fission yeast cell cycle is one example of a simulated biological system which contains many heterogeneous micro elements that perform specific functions in order to accomplish cell division. Applying the black-boxing framework reveals several macro local maxima that are stable throughout the biological sequence of the network model, and highlights the role of element Slp1 in stabilizing the cycle. Note that the typical approach of studying biological systems at a particular (macro) spatiotemporal scale is precisely to treat its next-lower level components as black boxes. Here we have proposed a theoretical framework to evaluate cause-effect power and the cause-effect properties of such a black-box system. If an organizational level corresponds to a local maximum of integrated information, then there will be cause-effect properties that emerge at that level, and there is knowledge to be gained by studying the system accordingly.

Finally, while local maxima reveal cause-effect properties to an investigator studying the system, the global maximum specifies the set of elements and spatiotemporal grain at which the system has most cause-effect power upon itself – from its own intrinsic perspective. According to integrated information theory, a set of elements at the spatial-temporal grain that defines the global maximum of intrinsic cause-effect power corresponds to a physical substrate of consciousness (Oizumi et al., 2014; Tononi, 2015).

---

## Supporting information Legends

**S1 Text. Full analysis of cause-effect power.** Detailed cause-effect structures for the examples presented in the main text.

**S2 Text. Black-boxing of degenerate or indeterministic systems.** Additional examples exploring the effects of degeneracy and indeterminism.

**S3 Text. The intrinsic perspective.** Additional discussion on analyzing cause-effect power from the intrinsic perspective.

# Black-boxing and cause-effect power:
# Supplementary Information

William Marshall[1], Larissa Albantakis[1], Giulio Tononi[1, *]

[1]*Department of Psychiatry, Center for Sleep and Consciousness, University of Wisconsin, Madison, WI, USA*
*Corresponding author: gtononi@wisc.edu*

## S1 Text – Full analysis of cause-effect power

Here we present a more detailed account of the integrated information analysis of the example systems discussed in the main text. All calculations were performed using the PyPhi software package in Python (Mayner et al., 2016).

For a given system in a specific state, the first step involves identifying its cause-effect structure, the set of mechanisms in the system. A mechanism is a set of elements that irreducibly constrains the past and future states of the system. Each member of the power set of system elements is tested as a potential mechanism. The set of system elements whose past states are most irreducibly constrained by the mechanism are its past purview (evaluated by the cause integrated information of the mechanism $\varphi_{cause}$). The set of system elements whose future states are most irreducibly constrained by the mechanism are the mechanism's future purview (evaluated by the effect integrated information $\varphi_{effect}$). The way that a mechanism, by being in its current state, constrains its purview elements is captured by its cause-effect repertoire, a pair of probability distributions over the past and future states of the purview elements (e.g. Fig. 3, main text). Note that these probabilities are obtained from the system's transition probability matrix (TPM) assuming a maximum entropy distribution for the marginal distribution of all possible past states. This corresponds to setting the system into all possible states with equal likelihood performing an interventionist causal analysis. $\varphi_{cause}$ and $\varphi_{effect}$ quantify the difference between the cause-effect repertoire, and the cause-effect repertoire under a partition of the mechanism as the earth-mover's distance between the two probability distributions (Oizumi et al., 2014). The overall integrated information of a mechanism is then the minimum of its $\varphi_{cause}$ and $\varphi_{effect}$. In sum, the complete specification of a mechanism thus includes its cause and effect purviews (the elements over which it has maximally irreducible power to constrain the past and future states), the cause-effect repertoires that specify those constraints, and its integrated information value ($\varphi$). The set of all mechanisms constitutes the system's cause-effect structure.

For a micro-level system, mechanisms are composed of micro elements and cannot include macro elements. Conversely, if a macro-level system is analyzed, only compositions of macro elements are tested and the potential causes and effects of individual micro elements within the black boxes are ignored. In other words, each level has a particular TPM, obtained from perturbing the system into all possible states with equal likelihood at that micro or macro level, which then determines the system's mechanisms at this particular level.

Next, to obtain the integrated information of the system, all possible directed partitions of the system are considered, to find the one that least affects the system's cause-effect structure. After each partition, the cause-effect structure is recalculated, and the result is compared to the cause-effect structure of the whole system. The partition that makes the least difference to the cause-effect structure is the minimum information partition (MIP), and the difference it makes, as measured using an extended earth movers distance (Oizumi et al., 2014), defines the integrated information of the system ($\Phi$). Note that, for $\Phi$, i.e. integrated information at the system level, all possible directed partitions of constituent micro elements are evaluated, regardless of whether the system is defined at a micro or macro level. This excludes the possibility to 'hide' micro elements without cause-effect power inside a black box, which would trivially increase the system's integration. In sum, the cause-effect structure (the set of all mechanisms) of a macro level system is evaluated purely at the macro level; its irreducibility, however, is evaluated by the partition between micro elements that makes the least difference to the macro cause-effect structure.

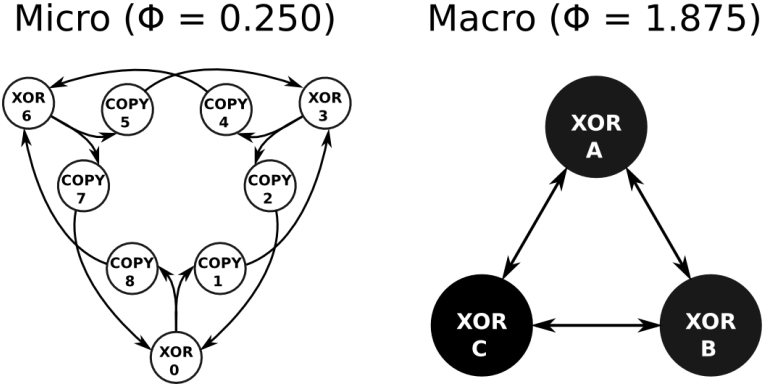

***Figure A****: Systems of micro elements with labels to facilitate description of the mechanisms. Left: Micro system with elements labeled 0-8. Right: Macro system with elements labeled A-C. All elements are in the 'OFF' state.*

## Example 1 – Composition and Integration

To describe the cause-effect structure of the micro system, we first assign labels to each of the micro elements in the system, as shown in Fig. A. There are three mechanisms in the cause-effect structure, they are all first order mechanisms and each one corresponds to an element implementing XOR logic.

| | Mechanism | Past Purview | Future Purview | φ |
|---|---|---|---|---|
| 0-8 unpartitioned | (0) | (2, 7) | (1, 8) | 0.5 |
| | (3) | (1, 5) | (2, 4) | 0.5 |
| | (6) | (4, 8) | (5, 7) | 0.5 |

To assess the integrated information Φ of the micro system, we identify the unidirectional system partition that makes the least difference to the cause-effect structure, termed the minimum information partition (MIP). For this system, the MIP is to cut all connections from (0, 1, 2, 3, 4, 5, 6, 7) to (8). Under this partition, the mechanism specified by element 0 is altered, its future purview is reduced from (1, 6) to only (1). Note that numbers in bold refer to mechanisms for which the partitioned cause-effect structure which is different from the unpartitioned cause-effect structure. Assessing the difference between the unpartitioned and partitioned cause-effect structure of the micro system using an extended earth-mover's distance (Oizumi et al., 2014), the resulting integrated information value of the micro system is Φ = 0.25.

| | Mechanism | Past Purview | Future Purview | φ |
|---|---|---|---|---|
| 0-8 partitioned | **(0)** | **(2, 7)** | **(1)** | **0.5** |
| | (3) | (1, 5) | (2, 4) | 0.5 |
| | (6) | (4, 8) | (5, 7) | 0.5 |

Next, we describe the cause-effect structure of the macro system displayed in Fig. A. The three black-box macro elements are constituted of micro elements A = (0, 2, 7), B = (1, 3, 4) and C = (5, 6, 8) with corresponding output elements (0), (4) and (6). This black-box system has three high-order mechanisms.

| | Mechanism | Past Purview | Future Purview | φ |
|---|---|---|---|---|
| ABC unpartitioned | (A, B) | (A, B, C) | (C) | 0.5 |
| | (A, C) | (A, B, C) | (B) | 0.5 |
| | (B, C) | (A, B, C) | (A) | 0.5 |

The MIP for this macro system cuts connections from (1, 3) to (0, 2, 4, 5, 6, 7, 8). After the partition, all of the mechanisms have been destroyed. Mechanisms (A, B) and (B, C) no longer have irreducible causes or effects, while the set of elements (A, C) has an effect but no cause. The integrated information of this system is $\Phi = 1.875$. Note that the MIP is a partition of micro elements; yet the black-box system has higher $\Phi$ because the partition affects the macro cause-effect structure more than it would affect the cause-effect structure of the corresponding micro system. This is because the mechanisms at the macro level are high-order mechanisms that constrain larger parts of the system (have larger purviews). These macro constraints are completely lost even under the micro partition that makes the least difference.

| | Mechanism | Past Purview | Future Purview | φ |
|---|---|---|---|---|
| ABC partitioned | **(A, B)** | **()** | **()** | **0** |
| | **(A, C)** | **()** | **(B)** | **0** |
| | **(B, C)** | **()** | **()** | **0** |

## Example 2 – Local Maxima

In this example with 55 elements we will not assign numbers to the elements. Instead, we will simply refer to each element based on the number of inputs and outputs it has, e.g., NOR(3, 1) for a NOR element with three inputs and one output. Each of the 55 elements specifies a first order mechanism, summarized in the table below.

| unpartitioned – micro | | | | |
|---|---|---|---|---|
| Multiplicity | Mechanism | Past Purview | Future Purview | φ |
| 30 | NOR(1, 1) | NOR(1, 6) | NOR(2, 1) | 0.25 |
| 15 | NOR(2, 1) | 2*NOR(1, 1) | NOR(3, 1) | 0.125 |
| 5 | NOR(3, 1) | 3*NOR(2, 1) | NOR(1, 6) | 0.5 |
| 5 | NOR(1, 6) | NOR(3, 1) | 6*NOR(1, 1) | 0.25 |

The minimum information partition (MIP) for this system is to cut the connections from a NOR(1, 1) element to the rest of the system. It doesn't matter which NOR(1, 1) element as they all have the same effect on their respective future purview. The result of the MIP is that one mechanism is destroyed, and another is altered. The integrated information of this system is $\Phi = 0.453$.

| partitioned – micro | | | | |
|---|---|---|---|---|
| Multiplicity | Mechanism | Past Purview | Future Purview | φ |
| 29 | NOR(1, 1) | NOR(1, 6) | NOR(2, 1) | 0.25 |
| **1** | **NOR(1, 1)** | **NOR(1, 6)** | **()** | **0** |
| 14 | NOR(2, 1) | 2*NOR(1, 1) | NOR(3, 1) | 0.125 |
| **1** | **NOR(2, 1)** | **NOR(1, 1)** | **NOR(3, 1)** | **0.125** |
| 5 | NOR(3, 1) | 3*NOR(2, 1) | NOR(1, 6) | 0.5 |
| 5 | NOR(1, 6) | NOR(3, 1) | 6*NOR(1, 1) | 0.25 |

One option for a macro system is to define black-box elements that implement AND and OR logic (Fig. 5 and 6, main text). This system has an average spatial grain size of 2.75. There is a symmetry in the system, so that the mechanisms specified by each OR gate are the same, and the mechanisms specified by each AND gate are also the same (the OR elements output to six AND elements and take inputs from three AND elements, while the AND elements all take inputs from two OR elements and output to one OR element). At this macro scale, the system has 15 black-box elements implementing AND logic and 5 black-box elements implementing OR logic, over two time steps, and each specifies a first order mechanism.

| unpartitioned – AND/OR black boxes | | | | |
|---|---|---|---|---|
| Multiplicity | Mechanism | Past Purview | Future Purview | φ |
| 5 | OR | 3*AND | 6*AND | 0.071 |
| 15 | AND | 2*OR | OR | 0.125 |

The MIP for this system is to cut the outputs of a NOR(2, 1) element that is one of the input elements of an AND black-box element. In this case, the mechanisms have the same cause-effect power, but one of the AND mechanisms and one of the OR mechanisms have reduced purviews, constraining less elements.

| partitioned – AND/OR black boxes | | | | |
|---|---|---|---|---|
| Multiplicity | Mechanism | Past Purview | Future Purview | φ |
| 4 | OR | 3*AND | 6*AND | 0.071 |
| **1** | **OR** | **3*AND** | **5*AND** | **0.071** |
| 14 | AND | 2*OR | OR | 0.125 |
| **1** | **AND** | **OR** | **OR** | **0.125** |

Another black-box system at a coarser macro scale has five black-box elements {A, B, C, D, E} over 4 time steps. Each black box implements a MAJORITY function over its three inputs, with a specialized connectivity pattern shown in Fig. 6, main text. Of the 31 ($2^N$-1) possible mechanisms from the power set of 5 elements, 30 specify irreducible past and future constraints:

| unpartitioned – MAJORITY black box | | | |
|---|---|---|---|
| Mechanism | Past Purview | Future Purview | φ |
| (A) | (C, D, E) | (B, C, D) | 0.25 |
| (B) | (A, D, E) | (C, D, E) | 0.25 |
| (C) | (A, B, E) | (A, D, E) | 0.25 |
| (D) | (A, B, C) | (A, B, E) | 0.25 |
| (E) | (B, C, D) | (A, B, C) | 0.25 |
| (A, B) | (A, C, E) | (C, D) | 0.2 |
| (A, C) | (A, B, C, D, E) | (A, B, C, D, E) | 0.2 |
| (A, D) | (A, B, C, D, E) | (A, B, C, D, E) | 0.2 |
| (A, E) | (B, D, E) | (B, C) | 0.2 |
| (B, C) | (B, D, E) | (D, E) | 0.2 |
| (B, D) | (A, B, C, D, E) | (A, B, C, D, E) | 0.2 |
| (B, E) | (A, B, C, D, E) | (A, B, C, D, E) | 0.2 |
| (C, D) | (B, C, E) | (A, E) | 0.2 |
| (C, E) | (A, B, C, D, E) | (A, B, C, D, E) | 0.2 |
| (D, E) | (A, C, D) | (A, B) | 0.2 |
| (A, B, C) | (A, B, C, D) | (A, B, C, D, E) | 0.2 |
| (A, B, D) | (A, C, E) | (B, C, D, E) | 0.257143 |
| (A, B, E) | (A, B, C, E) | (A, B, C, D, E) | 0.2 |
| (A, C, D) | (B, C, E) | (A, B, D, E) | 0.257143 |
| (A, C, E) | (B, D, E) | (A, B, C, D) | 0.257143 |
| (A, D, E) | (A, B, C, D) | (A, B, C, D, E) | 0.2 |
| (B, C, D) | (B, C, D, E) | (A, B, C, D, E) | 0.2 |
| (B, C, E) | (B, D, E) | (A, C, D, E) | 0.257143 |
| (B, D, E) | (A, C, D) | (A, B, C, E) | 0.257143 |
| (C, D, E) | (A, C, D, E) | (A, B, C, D, E) | 0.2 |

| (A, B, C, D) | (A, B, C, D, E) | (A, B, C, D, E) | 0.185709 |
| (A, B, C, E) | (A, B, C, D, E) | (A, B, C, D, E) | 0.185709 |
| (A, B, D, E) | (A, B, C, D, E) | (A, B, C, D, E) | 0.185709 |
| (A, C, D, E) | (A, B, C, D, E) | (A, B, C, D, E) | 0.185709 |
| (B, C, D, E) | (A, B, C, D, E) | (A, B, C, D, E) | 0.185709 |

The minimum information partition of this network is to cut the outputs of one of the NOR(1, 1) micro elements. By the symmetry in the system, there is an equivalent MIP in each of the black-box elements; however, due to the specialized connectivity structure, not all NOR(1, 1) elements are equivalent. One MIP option is to cut the hidden NOR(1, 1) micro element in black-box element A that receives input from D and outputs to the NOR(2, 1) micro elements along with the NOR(1, 1) micro element that receives input from C. As a result of the MIP, two of the mechanisms are destroyed (BCD and ABCD) and 15 others are modified (shown in bold), resulting in a $\Phi$ value of 2.333.

| partitioned – MAJORITY black box | | | |
|---|---|---|---|
| Mechanism | Past Purview | Future Purview | $\varphi$ |
| **(A)** | **(C, E)** | **(B, C, D)** | **0.25** |
| (B) | (A, D, E) | (C, D, E) | 0.25 |
| (C) | (A, B, E) | (A, D, E) | 0.25 |
| **(D)** | **(A, B, C)** | **(B, E)** | **0.25** |
| (E) | (B, C, D) | (A, B, C) | 0.25 |
| **(A, B)** | **(D, E)** | **(C, D)** | **0.227273** |
| **(A, C)** | **(A, B, C, E)** | **(A, B, C, D, E)** | **0.2** |
| **(A, D)** | **(A, B, C, E)** | **(B, C, D, E)** | **0.2** |
| **(A, E)** | **(B, C, E)** | **(B, C)** | **0.2** |
| (B, C) | (B, D, E) | (D, E) | 0.2 |
| **(B, D)** | **(A, B, C, D, E)** | **(B, C, D, E)** | **0.2** |
| (B, E) | (A, B, C, D, E) | (A, B, C, D, E) | 0.2 |
| **(C, D)** | **(B, C, E)** | **(A, B, D, E)** | **0.2** |
| (C, E) | (A, B, C, D, E) | (A, B, C, D, E) | 0.2 |
| (D, E) | (A, C, D) | (A, B) | 0.2 |
| **(A, B, C)** | **(A, B, C, D)** | **(A, B, C, D, E)** | **0.181816** |
| (A, B, D) | (A, C, E) | (B, C, D, E) | 0.257143 |
| (A, B, E) | (A, B, C, E) | (A, B, C, D, E) | 0.2 |
| **(A, C, D)** | **(B, C, E)** | **(B, D, E)** | **0.257143** |
| **(A, C, E)** | **(B, C, E)** | **(A, B, C, D)** | **0.257143** |
| **(A, D, E)** | **(A, B, C, D, E)** | **(A, B, C, D, E)** | **0.142158** |
| (B, C, E) | (B, D, E) | (A, C, D, E) | 0.257143 |
| (B, D, E) | (A, C, D) | (A, B, C, E) | 0.257143 |
| (C, D, E) | (A, C, D, E) | (A, B, C, D, E) | 0.2 |
| **(A, B, C, E)** | **(A, B, C, D, E)** | **(A, B, C, D, E)** | **0.207691** |
| **(A, B, D, E)** | **(A, B, C, D, E)** | **(A, B, C, D, E)** | **0.25** |
| **(A, C, D, E)** | **(A, B, C, D, E)** | **(A, B, C, D, E)** | **0.161903** |
| (B, C, D, E) | (A, B, C, D, E) | (A, B, C, D, E) | 0.185709 |

# Black-boxing and cause-effect power: Supplementary Information

William Marshall[1], Larissa Albantakis[1], Giulio Tononi[1, *]

[1]*Department of Psychiatry, Center for Sleep and Consciousness, University of Wisconsin, Madison, WI, USA*
*Corresponding author: gtononi@wisc.edu*

## S2 Text – Black-boxing of degenerate or indeterministic systems

As demonstrated in (Hoel et al., 2013, 2016), the main factor enabling an increase in intrinsic cause-effect power through coarse-graining is a reduction in indeterminism and degeneracy at the macro level. An increase in intrinsic cause-effect power by reducing degeneracy is also possible through black-boxing, as shown below ("Degeneracy example"). However, black-boxing may increase indeterminism (see below "Propagation delay through noisy channel").

## Propagation delay through noisy channels

In Example 1 (main text), we demonstrated that black-boxing a system with deterministic propagation delay may lead to an increase in intrinsic cause-effect power, as it can increase system integration through emergent high-order mechanisms. However, indeterminism in the system may influence whether a black-boxed macro level can "beat" the micro level in terms of cause-effect power.

Here we explore the effect of noise on black-boxing by considering a scenario in which propagation delays occur over noisy channels. In this case, the COPY elements take a single input and then output the same value with probability $p$ in [0.5, 1]. The original results of Fig. 2 (main text) refer to a noiseless channel ($p = 1$). A completely noisy channel ($p = 0.5$) would have no cause-effect power at any level (macro or micro). The integrated information analysis is performed on both the micro and black-box physical system for several different values of $p$ (see Table 1). The number and orders of mechanisms is the same for all values of $p > 0.5$, but the $\varphi$ value of each mechanism decreases as $p$ decreases, and it does so more steeply for the black-box system than for the micro system. Along with $\varphi$, the overall integrated information $\Phi$ of both the black-box and micro systems decrease as the amount of noise increases. Since $\Phi$ of the black-box system declines faster with increasing noise than $\Phi$ of the micro system, the macro system "beats" the micro for $p > 0.6$, whereas the micro level system "wins" when $p \le 0.6$. The reduction in cause-effect power due to indeterminism can outweigh the increase due to high-order mechanisms, since indeterminism (noise) disproportionately affects high-order mechanisms. Therefore, several interrelated aspects of cause-effect power have to be considered when assessing if the "macro beats the micro," including the presence of high-order mechanisms, indeterminism, and degeneracy.

**Table 1:** *Integrated information for various noise levels (p) in the propagation delay network (Fig. 3 main text)*

| $p$ | 1.0 | 0.9 | 0.8 | 0.7 | 0.6 | 0.5 |
|---|---|---|---|---|---|---|
| Micro $\Phi$ | 0.250 | 0.160 | 0.090 | 0.040 | 0.010 | 0.000 |
| Micro $\varphi$ | 0.5 | 0.4 | 0.3 | 0.2 | 0.1 | 0 |
| Black box $\Phi$ | 1.875 | 1.046 | 0.363 | 0.070 | 0.004 | 0.000 |
| Black box $\varphi$ | 0.500 | 0.320 | 0.180 | 0.080 | 0.020 | 0.000 |

# Degeneracy example

Reducing degeneracy, the convergence of multiple past system states onto the same current system state, is one way in which a coarse-grained macro level can achieve higher cause-effect power than its corresponding micro level (Hoel et al., 2013, 2016). Here we show that black-boxing micro elements into macro elements can also be exploited to counteract degeneracy at the micro level.

Fig. A shows a system of six micro elements – four COPY gates and two AND gates that are black-boxed into two macro elements, which correspond to macro COPY gates. As illustrated by the TPM in Fig. A (left, bottom), the micro level has a high degree of degeneracy (*i.e.* multiple rows leading to the same column in the TPM). In the example, two COPY micro elements input to a single AND micro element, whose current state is OFF. This implies degeneracy in the system since three states of the COPY elements (OFF, OFF), (ON, OFF) and (OFF, ON) all lead to the same state of the AND element (OFF). The micro system in state 'all OFF' specifies six first order mechanisms; the AND elements specify mechanisms with $\varphi = 0.167$; and the COPY elements specify mechanisms with $\varphi = 0.25$. There are no high-order mechanisms. The integrated information for the micro system is $\Phi = 0.215$.

We then consider the system at the macro level, after black-boxing each AND micro element with its two inputting COPY micro elements. Each black-box macro element turns out to implement, over two time steps, a macro COPY logic, specifying a first-order mechanism with $\varphi = 0.5$. Again, there are no high-order mechanisms. However, the macro system has no degeneracy, since no two past states converge onto the same current state (see macro TPM in Fig. A). For this reason, the integrated information of a COPY element in the macro system is higher ($\varphi = 0.5$) than in the micro system ($\varphi = 0.25$), and so is the overall integrated information for the black-box macro system ($\Phi = 0.639$). Thus, appropriately black-boxing micro into macro elements reduces degeneracy and in doing so increases the cause-effect power of the system, as measured by integrated information $\Phi$.

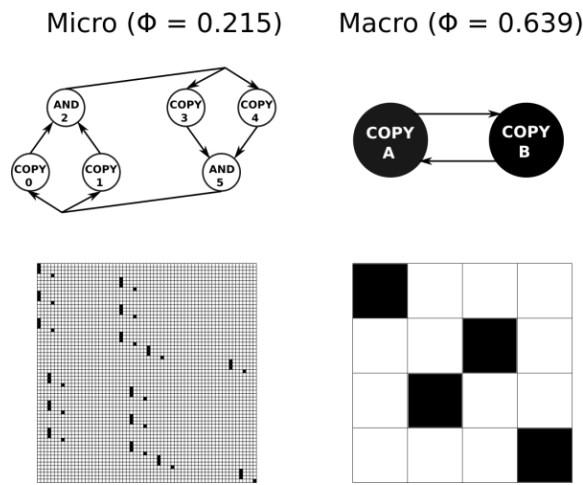

Micro ($\Phi = 0.215$)     Macro ($\Phi = 0.639$)

***Figure A:*** *Left: Two AND elements that each receive inputs from two COPY elements and send output to two other COPY elements, along with the TPM calculated from systematic perturbation of the elements. The current state of all micro elements is OFF. The integrated information of the micro system is $\Phi = 0.215$. Right: Black-box elements consisting of AND elements as outputs, and the corresponding COPY elements hidden within. The current macro state of the black-box elements is OFF, corresponding to the current micro state of the micro elements that define their output. The TPM for this system is found by perturbing the inputs to the black-box element in all possible ways, it is determined that it implements COPY logic over two time steps. The black-box macro system has $\Phi = 0.639$.*

To describe the cause-effect structure of the micro system in detail, we use the labels shown in Fig. A.

| | Mechanism | Past Purview | Future Purview | φ |
|---|---|---|---|---|
| micro unpartitioned | (0) | (5) | (2) | 0.25 |
| | (1) | (5) | (2) | 0.25 |
| | (2) | (0, 1) | (3, 4) | 0.167 |
| | (3) | (2) | (5) | 0.25 |
| | (4) | (2) | (5) | 0.25 |
| | (5) | (3, 4) | (0, 1) | 0.167 |

For this system, the MIP is to cut all connections from (1) to (0, 2, 3, 4, 5). Note that there are other equivalent MIPs; we focus on this specific cut without loss of generality. Under the partition, element 1 no longer has an effect and thus does not specify a mechanism, and the mechanism specified by element 2 is altered, as its past purview is reduced from (0, 1) to only (0). Comparing the unpartitioned and partitioned cause-effect structure, we find that the resulting integrated information value for the micro system is $\Phi = 0.215$. Entries in bold highlight mechanisms in the partitioned cause-effect structure that are different from those in the unpartitioned cause-effect structure.

| | Mechanism | Past Purview | Future Purview | φ |
|---|---|---|---|---|
| micro partitioned | (0) | (5) | (2) | 0.25 |
| | **(1)** | **(5)** | **()** | **0** |
| | **(2)** | **(0)** | **(3, 4)** | **0.167** |
| | (3) | (2) | (5) | 0.25 |
| | (4) | (2) | (5) | 0.25 |
| | (5) | (3, 4) | (0, 1) | 0.167 |

We then consider a black-box system of two COPY elements over two time steps. The black-box elements are constituted of micro elements A = (0, 1, 2) and B = (3, 4, 5), with corresponding output elements 2 and 5. The cause-effect structure of the black-box system is:

| | Mechanism | Past Purview | Future Purview | φ |
|---|---|---|---|---|
| black box unpartitioned | A | B | B | 0.5 |
| | B | A | A | 0.5 |

The MIP for this system is to cut all connections from (0) to (1, 2, 3, 4, 5), that is, cut the outputs of micro element 0 (by symmetry, cutting the outputs of elements 1, 3 or 4 would be equivalent). After the partition, both mechanisms have had their irreducible cause-effect power diminished, and the result is $\Phi = 0.639$.

| | Mechanism | Past Purview | Future Purview | φ |
|---|---|---|---|---|
| black box partitioned | **A** | **B** | **B** | **0.167** |
| | **B** | **A** | **A** | **0.25** |

# Black-boxing and cause-effect power:
# Supplementary Information

William Marshall[1], Larissa Albantakis[1], Giulio Tononi[1, *]

[1]*Department of Psychiatry, Center for Sleep and Consciousness, University of Wisconsin, Madison, WI, USA*
*Corresponding author: gtononi@wisc.edu*

## S3 Text – The intrinsic perspective

The procedure for calculating the intrinsic cause-effect power of systems at a micro level is documented in detail in previous work (Oizumi et al., 2014, Mayner et al., 2016). Here we discuss aspects that deserve further comments when the analysis of intrinsic cause-effect power is applied to macro systems.

## Intrinsic cause-effect power

To perform the intrinsic causal analysis of a system, its elements are perturbed with equal probability into all possible states and the resulting state transitions are observed. As with other interventional accounts of causality, by *setting* the elements into different states rather than merely observing them, true causal relationships can be distinguished from correlations. Additionally, perturbing the system into all possible states allows all counterfactuals to be evaluated, to assess the specificity of the causal relationship between elements and quantify cause-effect power. The goal of this exhaustive perturbational analysis is to unfold the full intrinsic cause-effect power of the system, that is, how the current state of system elements constrains their potential past and future states. Importantly, when analyzing the system at the micro level the micro elements of the system are set into all their possible micro states with equal probability. The same is done when analyzing the system at the macro level, except that this time it is the macro elements that are set into all their possible macro states with equal probability. In general, equal probability for macro states does *not* correspond to a uniform distribution of micro states.

When performing intrinsic causal analysis on a black-box system, some specific considerations have to be taken into account. Most importantly, one must capture the constraints due to the macro state of the black box itself (its output) but not any constraints due to the state of the micro elements within the black box. In this regard, it is useful to consider micro elements that fall into three different categories; micro elements that impose constraints within their corresponding macro element (Fig A-A), micro elements that impose constraints on other black boxes (Fig A-B; connection from F to D), and micro elements that are outside the system (Fig A-C; D is outside the system).

A single macro state may be consistent with multiple micro states of its micro elements, which nevertheless may impose different constraints on the past/future states of the system. Such micro constraints that do not originate from a black box's output element at the relevant macro update must be discounted throughout the macro causal analysis and are treated as noise. For example, consider the black-box element γ in Fig. A-A, which has micro elements C and D as inputs (receiving from the output elements of black boxes α and β respectively) and E as its output. There is feedback within this black box (bidirectional connection between micro elements C and D), hence the state of micro elements C and D matters when specifying the input-output relationship of the black box (see Table 2). However, the state of constituent micro elements should not contribute to the cause-effect power of a macro system. Therefore, when perturbing the system into a particular macro state, the relevant micro elements are perturbed with equal probability into all possible micro states that are consistent with the macro state. In practice, this procedure amounts to averaging across the input-output relation for all possible states of the hidden micro elements within the black box, resulting in the truth table shown in Fig A-A, which is the average of the four truth tables resulting from the four possible states of CD (00, 10, 01, 11) (Table 2). Another instance (not dealt with in the current work) is when the output of a generalized macro element (black box) is defined as a coarse-graining of micro elements. In this case, the specific identity of these micro constituents should not contribute to the intrinsic cause-effect power of the macro system. However, in certain situations, such as when a single micro element from one coarse-grained macro element provides output to two different macro elements, constraints due to specific micro constituents can manifest in the macro TPM as "instantaneous causation" or "conditional dependence" (i.e., the current state of a macro element constrains the current state of other macro elements) and must be discounted in the causal analysis.

Another type of micro constraint that needs to be discounted while evaluating the intrinsic cause-effect power of a macro-level system is 'lateral' connections between black boxes, i.e. when a micro element hidden within a black box constrains a micro element within a different black box. To ensure that only constraints due to the state of macro elements are captured in the analysis, all connections originating from micro elements within a black box and terminating outside of it are injected with noise. This includes not only connections belonging to micro elements hidden within the black box, but also the micro output element of the black box at times other than the macro time step being considered. For example, in Fig. A-B, the connection from micro element F to micro element D must be noised, since F is not an output element of a black box. Similarly, when assessing whether black-box elements α and β constrain the future state of δ over 4 time steps, the output element E of γ must be noised throughout the macro update. This is to ensure that constraints due to the state of micro elements within γ are not counted towards the cause-effect power of α or β, which here do not constrain δ directly. The black-box element γ can be thought as "screening off" the indirect effect of α and β on δ, because α and β can only effect δ via the intermediate element γ.

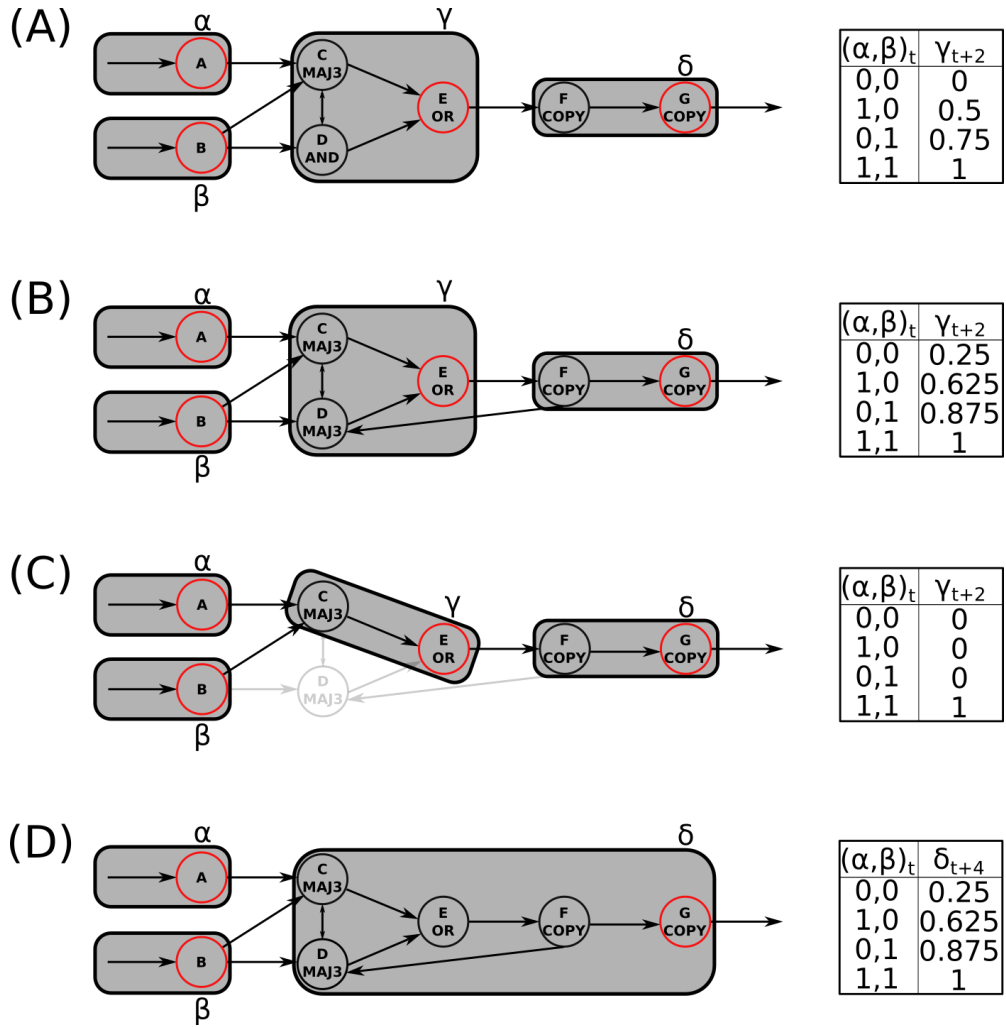

**Figure A:** *Examples of partial systems of micro elements (A, B, C, D, E, F, G) with several different black-boxings into macro elements and the corresponding input-output function. (A): Four black-box elements α, β, γ, δ. Elements α and β have a joint constraint on γ, but it is not fully specific (there is indeterminism). All outputs of hidden micro elements are internal to their corresponding black box. (B): Similar to the (A) except constituent element F has an output that leaves its black box. The effect of F on D is not intrinsic to the macro system and must be noised during the entire causal analysis. As a result, the effects on gamma are less deterministic. (C): A similar black-boxing as in (B), except micro element D is outside the system, rather than within γ. The element D in its current state 0 is thus taken as a background condition. In this situation, the effect of α and β on γ is more specific. (D): A potential black-boxing with only three elements. In panels (A), (B) and (C), the effect of α and β on δ is screened-off by γ. In this case, the micro connection from E to F is within a black box rather than between so α and β have a direct effect on δ (it is no longer screened-off).*

*Table 2: Input-output relation for different initial states of hidden micro elements (C, D) of black box γ in panel (A) of Fig. A.*

| If (C, D)$_t$ = (0, 0) | | |
|---|---|---|
| (A, B)$_t$ | (C, D)$_{t+1}$ | E$_{t+2}$ |
| 0, 0 | 0, 0 | 0 |
| 1, 0 | 0, 0 | 0 |
| 0, 1 | 0, 0 | 0 |
| 1, 1 | 1, 0 | 1 |

| If (C, D)$_t$ = (1, 0) | | |
|---|---|---|
| (A, B)$_t$ | (C, D)$_{t+1}$ | E$_{t+2}$ |
| 0, 0 | 0, 0 | 0 |
| 1, 0 | 0, 0 | 0 |
| 0, 1 | 0, 1 | 1 |
| 1, 1 | 1, 1 | 1 |

| If (C, D)$_t$ = (0, 1) | | |
|---|---|---|
| (A, B)$_t$ | (C, D)$_{t+1}$ | E$_{t+2}$ |
| 0, 0 | 0, 0 | 0 |
| 1, 0 | 1, 0 | 1 |
| 0, 1 | 1, 0 | 1 |
| 1, 1 | 1, 0 | 1 |

| If (C, D)$_t$ = (1, 1) | | |
|---|---|---|
| (A, B)$_t$ | (C, D)$_{t+1}$ | E$_{t+2}$ |
| 0, 0 | 0, 0 | 0 |
| 1, 0 | 1, 0 | 1 |
| 0, 1 | 1, 1 | 1 |
| 1, 1 | 1, 1 | 1 |

In Fig. A-B, both micro elements C and D constrain the output element E, hence C and D would naturally seem to belong inside the black box γ. Nevertheless, the search for local maxima of intrinsic cause-effect power must consider all alternatives, including one in which D is taken to be a background condition rather than a hidden element within γ (or part of any other black box). In this alternate system (Fig. A-C), the state of D (OFF) is fixed as a background condition. In this case, the input-output relation for black box γ changes to the one shown in Fig A-C. From the figure alone, one cannot determine which of the two systems (top or middle panel) should qualify as the local maximum. However, it is apparent that the effect of α and β on γ is not fully deterministic when D is hidden inside γ, but it becomes deterministic when D is treated as a background condition. Furthermore, the repertoire of possible past states of γ is more degenerate when D is hidden inside γ as compared to when D is treated as a background condition (in this case there is only one possible past state of γ = 1, while with D inside the black box all four states of α and β could have led to γ = 1 with some probability). Thus, everything else being equal, the system with D as a background condition is more deterministic and less degenerate than the system with D hidden inside γ and should therefore have higher Φ. This result suggests that even if, from an extrinsic perspective, a set of micro elements may appear to constitute a macro element, from the intrinsic perspective only the set of micro elements that contribute to maximizing cause-effect power (i.e., the "skeleton" mediating the strongest constraints) actually constitutes the macro element.

## Integration and Exclusion

When evaluating cause-effect power at macro scales, we have to consider the micro elements that *constitute* the macro-level system. As stated in the main text (section "Black-boxing" parts iii and iv), the integration and exclusion postulates both apply to the constituents of black-box systems. By the integration requirement, the set of constituents must be irreducible; two unrelated systems cannot be black-boxed together since their constituents are not integrated (Fig. B-A, B-B). Moreover, micro input (or output) elements cannot be black-boxed (Fig. B-C) because they lack causes (or effects) within the system. Only when the set of constituents is integrated (Fig. B-D) can a macro system be integrated.

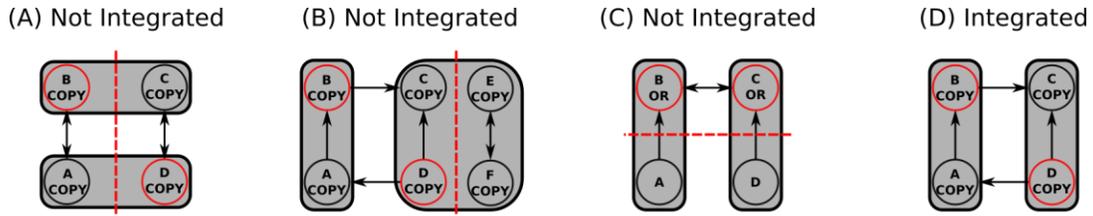

***Figure B***: *Four examples of potential black-box systems and their constituent micro elements. The constituents of a macro system must be irreducible (integration). Among the examples in the figure only (D) has a properly integrated set of constituents, while (A), (B) and (C) are reducible, with the corresponding cut drawn as a dashed red line. (A, B) Two systems that are not integrated at the micro level cannot constitute a macro, black-boxed system; (C) Elements that provide only inputs to (or only outputs from) the system cannot be constituents of a black box.*

By the exclusion postulate, a (macro) element must be definite. Thus, a micro constituent cannot contribute to multiple black boxes within a system (Fig. C).

## No Blackbox Overlap

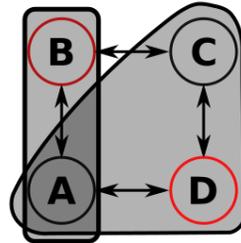

***Figure C***: *Two examples of micro constituents having their cause-effect power double counted. (A) The micro element A is contributing to two different black-box elements (A, B) and (A, C, D). The exclusion postulate rules out this potential black-box system.*